\documentstyle[12pt]{article}
\begin{document}

\begin{center}
{\Large {\bf Double exchange and orbital correlations in electron doped 
manganites}}\\
\vspace{1.0cm} 
{{\bf Tulika Maitra$^{*}$}\footnote{email: tulika@phy.iitkgp.ernet.in}  
and {\bf A. Taraphder}$^{*\dagger}$\footnote{email: arghya@phy.iitkgp.ernet.in,
arghya@pa.msu.edu}}\\ 

\noindent $^{*}$Department of Physics \& Meteorology and  
Centre for Theoretical Studies,\\
Indian Institute of Technology, Kharagpur 721302 India \\   
and\\
$^{\dagger}$Department of Physics \& Astronomy, Michigan State University,\\
East Lansing, MI 48823-1116 
\end{center}
\begin{abstract}
\vspace{.5cm} 
A double exchange model for degenerate $e_g$ orbitals with intra- and 
inter-orbital interactions has been studied for the electron doped manganites
A$_{1-x}$B$_{x}$MnO$_3$ ($x > 0.5$). We show that such a model reproduces
the observed phase diagram and orbital ordering in the intermediate bandwidth
regime and the Jahn-Teller effect, considered to be crucial for the region
$x<0.5$, does not play a major role in this region. Brink and Khomskii have 
already pointed this out and stressed the relevance of the anistropic hopping
across the degenerate $e_g$ orbitals in the infinite Hund's coupling limit.
From a more realistic calculation with finite Hund's coupling, we show that
inclusion of interactions stabilizes the C-phase, the antiferromagnetic
metallic A-phase moves closer to $x=0.5$ while the ferromagnetic phase shrinks.
This is in agreement with the recent observations of Kajimoto et. 
al. and Akimoto et. al.
\end{abstract}
\vspace{.3cm} 

The perovskite manganites have been at the centre of attention\cite{sdm,imada}
recently as systems like Pr$_{1-x}$Ca$_x$MnO$_3$, La$_{1-x}$Ca$_x$MnO$_3$, 
exhibit colossal magnetoresistance (CMR). These two extensively studied 
systems have relatively low bandwidths. The observation of CMR even in the
wider bandwidth materials like Pr$_{1-x}$Sr$_x$MnO$_3$\cite{tomi95} and
Nd$_{1-x}$Sr$_x$MnO$_3$ for $x\ge 0.5$\cite{kuwa99} have prompted a series
of careful experimental work on the magnetic phase diagram of all of these
systems particularly in the region $x \ge 0.5$ (the so called {\em electron
doped regime}). This region shows a rich variety of magnetic
and orbital ordering and only recently the systematics of the phase diagram
with externally controlled bandwidth have begun to emerge\cite{akimoto98}.

In their study of Pr$_{1-x}$Sr$_x$MnO$_3$, Kajimoto et. al.\cite{kajimoto01}
have summarized the nature of magnetic ordering for a series of manganites
having different bandwidths (Fig.1 in ref.\cite{kajimoto01}). They observe
that there is no CE phase in many of these systems and the sequence of phases
in the electron doped region for moderate to large bandwidth systems follows
the order ferromagnetic (F) $\rightarrow$ A-type antiferromagnetic (AFM) 
$\rightarrow$
C-type AFM and finally to G-type AFM phase. The F phase close to
$x=0.5$ is very narrow and survives for systems with bandwidths above 
Pr$_{0.5}$Sr$_{0.5}$MnO$_3$ (e.g., in (La$_{0.5}$Nd$_{0.5}$)$_{1-x}$Sr$_x$
MnO$_3$ there is a small sliver of F metallic phase\cite{akimoto98} ). 
A-phase exists in a small region while the C-phase covers
the widest region in the phase diagram. The general trend is
that with decreasing bandwidth the F-phase reduces, A-phase moves closer
to $x=0.5$ while the C-phase grows. It is also observed that the
gradual building of AFM correlations, starting from $x=0.5$, is preempted by
the orbital ordering in the A and C phases\cite{kajimoto99,yoshi}. There does
not seem to be any convincing evidence in favour of phase separation in this
region\cite{kajimoto01}.

In the absence of Jahn-Teller (JT) splitting the two e$_g$ orbitals of Mn ion
are degenerate. The doped manganite R$_{1-x}$A$_{x}$MnO$_{3}$
has $y=1-x$ number of electrons in the e$_g$ orbitals and the filling,
therefore, is $\frac{y}{4}$. In the foregoing, we restrict ourselves to the
region $x\ge 0.5$ i.e., $y \le 0.5$. At the $x=1$ end, the band is empty
and the physics is governed entirely by the exchange between the t$_{2g}$
electrons in the neighbouring sites. On doping, the kinetic energy of
electrons in the e$_g$ levels begin to compete with the AF 
superexchange (SE) between neighbouring t$_{2g}$ spins via Hund's coupling 
and this leads to a rich variety of magnetic and orbital structures.
A model incorporating this physics has recently been proposed
by Brink and Khomskii\cite{khom} (hereinafter referred to as BK). 

The model BK used for the electron-doped manganites contains three terms

$$ H = J_{AF} \sum_{<ij>} {\bf S_{i}}.{\bf S_{j}} - J_{H} \sum_{i} {\bf S_{i}}.
{\bf s_{i}} - \sum_{<ij>\sigma,\alpha,\beta}t_{i,j}^{\alpha \beta} 
c_{i,\alpha,\sigma}^{\dagger} c_{j,\beta,\sigma}\eqno(1)$$

The first term represents the AF exchange between t$_{2g}$ spins, the second
term is the Hund's coupling between t$_{2g}$ and e$_g$ spins at each site and
the third one provides hopping between the two orbitals\cite{khom73}
($\alpha,\beta$ take values 1 and 2 for d$_{x^2-y^2}$ and d$_{3z^2-r^2}$
orbitals, corresponding to the choice $\xi_i=0$ in Ref.\cite{hotta00}).

The existence of the degenerate e$_g$ orbitals with the asymmetric hopping
integrals t$_{ij}^{\alpha \beta}$ makes (1) very different from the usual
DE model\cite{sdm}. BK treated the t$_{2g}$ spins quasi-classically and $J_H$
was set to infinity. Canting in the x-z plane was included through the
effective hoppings $t_{xy}=tcos(\theta_{xy}/2)$ and $t_z=tcos(\theta_z/2).$ 
Here $\theta_{xy}$ ($\theta_z$) is the angle between nearest neighbour
t$_{2g}$ spins in the x-y plane (z-direction). The superexchange energy per
state, then, is $E_{SE}=\frac
{J_{AF}S_0^2}{2}(2cos\theta_{xy}+cos\theta_z).$ At this level of approximation,
the problem reduces to solving the 2$\times$2 matrix equation 
$|| t_{\alpha\beta}-\epsilon \delta_{\alpha\beta} || =0$ for a system of
spinless fermions and minimizing the total energy with respect to $\theta_{xy}$
and $\theta_z.$ In the uncanted state, $\theta_{xy}=\theta_z=0$ implies
F phase, $\theta_{xy}=\theta_z=\pi$ G-type, $\theta_{xy}=\pi$ and
$\theta_z=0$ C-type and $\theta_{xy}=0$ and $\theta_z=\pi$ A-type
AFM phases.  

Remarkably, the phase diagram obtained by BK based on such simplifying
assumptions indeed shows the different magnetic phases seen experimentally in 
Nd$_{1-x}$Sr$_{x}$MnO$_{3}$ and Pr$_{1-x}$Sr$_{x}$MnO$_{3}$ in the region $x 
> 0.5$ although the G-phase, expected for
the nearly empty band (close to $x=1$) and the F state at $J_{AF}
\rightarrow 0$ were, however, not recovered. The limit of
infinite Hund's coupling which BK worked with is somewhat unphysical
for the manganites considered\cite{dagrev,coey,misra}. In a more realistic
treatment Pai\cite{pai} considered the limit of finite J$_H$ and succeeded
in recovering the G and F phases. 

From these results BK argue that the degeneracy of $e_g$ orbitals and the
anisotropy of hopping are crucial and the JT effect not quite as relevant
since the number of JT
centres is low in the range of doping considered. The effect of disorder,
completely ignored in this model, does not seem to play a major role in the
magnetic phase diagram\cite{pai}. 

Neither of the treatments of BK and Pai include the interactions present in
the system, namely the inter- and intra-orbital Coulomb interactions as well
as the intersite Coulomb interaction\cite{dagrev,coey,cuoco}. Although
for low doping the interactions are expected to be ineffective, with increase
in doping they tend to localize the carriers and preferentially enhance the
orbital ordering. This affects the F-phase and alters the relative
stability of A and C phases. It is, therefore, necessary to include them in the
Hamiltonian (1) and look for their effects on the phase diagram.
A natural extension to the model (1) is then\cite{dagrev} 

$$ H_{int} =U\sum_{i\alpha}n_{i\alpha\uparrow}n_{i\alpha\downarrow}+U^{\prime}
\sum_{i\sigma\sigma^{\prime}}n_{i1\sigma}n_{i2\sigma^{\prime}}\eqno(2)$$

Here $U$ and $U^{\prime}$ represent the intra- and inter-orbital Coulomb
interaction strengths. For the systems concerned, we are not looking for
the charge ordered states\cite{comm_CO} and neglect longer range interactions. 
We take the interactions $U$ and $U^\prime$ as parameters
as in\cite{hotta00,misra} while in reality, these are related through
the Racah parameters(see \cite{dagrev} and references therein).
We treat the spin system quasi-classically, but unlike BK we work at
finite Hund's coupling. In the uncanted states, we assume for 
the t$_{2g}$ spin ${\bf S_{i}}={\bf S}_{0} \exp(i{\bf Q.r_{i}}),$ 
where the choice of ${\bf Q}$ determines different spin arrangements for 
the core spins. In the infinite J$_H$ limit, the e$_g$ electron spins
would be forced to follow the core spins leading to the freezing of their
spin degrees of freedom. A finite value for J$_H$, however, 
allows for fluctuations and the spin degrees of freedom, along
with anisotropic hopping across the two orbitals, play a crucial role. 

Let us first look at the Hamiltonian (1) i.e., set  $U = U^{\prime} =0 $.
For t$_{2g}$ spin configurations described above, it reduces to 
$$ H = \sum_{{\bf k},\alpha,\beta,\sigma} \epsilon_{\bf k}^{\alpha\beta}
c_{{\bf k}\alpha\sigma}^{\dagger} c_{{\bf k}\beta\sigma}
-J_HS_0 \sum_{{\bf k},\alpha} c_{{\bf k}\alpha\uparrow}^{\dagger} 
c_{{\bf k+Q}\alpha\uparrow}+J_HS_0 \sum_{{\bf k},\alpha} c_{{\bf k}\alpha
\downarrow}^{\dagger} c_{{\bf k+Q}\alpha\downarrow}\eqno(3) $$
\noindent where we have followed the notation in\cite{khom73} for 
$\epsilon_{\bf k}^{\alpha \beta}$. 

We calculate the ground state energy by diagonalization of the above
Hamiltonian in a finite momentum grid (numerical results converged by a
grid size 64$\times 64\times 64$) as a function
of $J_{H}$ for a range of values of $J_{AF}$. The magnetic phase diagram for
J$_{AF}S_{0}^{2}=0.05$ in the $J_H-x$ plane is shown in Fig.1a. The phase
diagram in J$_{AF}-x$ plane for J$_HS_{0}=$10 is plotted in Fig.2a. All
energies are measured in units of the hopping $t$. There is no
general agreement on the values of the parameters involved\cite{dagrev}. From
photoemission and optical studies\cite{dagrev,coey}
and LDA analysis\cite{satpat}) one can glean a range of typical
values $0.1eV < t < 0.3eV$, J$_H\simeq 1.5-2$ eV and   
J$_{AF}\simeq 0.03t-0.1 t$ (Maezono et. al.\cite{maezono} quote
a lesser value of J$_{AF}=0.01t$). 

At the $x=1$ end, with empty e$_g$ orbitals, the only
contribution to energy comes from the SE interaction leading to the G-type
AFM phase. On doping by electrons the C-phase appears first with
orbital ordering (of the d$_{z^2}$ orbitals) along the z-direction. The
stability of the A-phase comes from the ordering of d$_{x^2-y^2}$ orbitals in
the xy plane. The gain in kinetic energy due to the planar and one dimensional
orbital order, induced by the anisotropic hopping integral, more than 
offsets the loss of SE energy. The orbital order drives the corresponding 
magnetic order as well - in the  A-phase the spins have planar FM order 
and AFM order along the z-direction whereas in the C-phase it is reversed.
The 3D magnetically ordered F- and G-phases show no orbital ordering.
The phase transitions are therefore characterised by the density of
states (DOS) reflecting the underlying 1, 2 and 3 dimensional characters of 
the different phases. The phase diagrams are shown
in Figs.1a and 2a, for typical values of the parameters J$_HS_{0}=10$ and
J$_{AF}S_{0}^{2}=0.05$. The sequence of  G, C, A and finally the F 
phase with complete alignment of spins is observed. 

In the infinite Hund's coupling limit the ``wrong" spin sector of the Hilbert
space was projected out by BK. When the system is doped, spin canting is
the only channel for delocalization of doped carriers, albeit with a loss in
SE energy. At finite J$_H$, however, the wrong spins are no longer
as ``costly" and canting is expected to reduce. Canting is included through
the choice ${\bf S_{i}}={\bf S}_{0}(sin\theta_{i},0,cos\theta_{i}).$ The 
Hund's coupling term in the Hamiltonian becomes $ H_{hund}
=-J_HS_0\sum_{i,\alpha}cos\theta_i(c_{i\alpha\uparrow}^{\dagger}c_{i\alpha
\uparrow}-c_{i\alpha\downarrow}^{\dagger}c_{i\alpha\downarrow})-J_HS_0\sum_{i,
\alpha}sin\theta_i(c_{i\alpha\uparrow}^{\dagger}c_{i\alpha\downarrow}+
c_{i\alpha\downarrow}^{\dagger}c_{i\alpha\uparrow}).$ In this case the 
different magnetic phases need to be defined at the outset. The convention 
used by BK to define the magnetic phases are: it is A-type when
$\theta_{xy} < \theta_z$ and C-type when $\theta_{xy} > \theta_z.$ In the
canted G and F phases $\theta_{xy}$ and $\theta_z$ are close to $180^0$ and 
$0^0$ respectively, although, it is obvious that the canted G-phase and A-phase
are synonymous in a certain region. However, orbital order can be used to
delineate the two phases\cite{comm_BK}.

Proceeding as before, the ground state energy for different $\theta_{xy}$ and 
$\theta_z$ is obtained. The qualitative phase diagram is very similar to the
uncanted case except for small shifts in the phase boundaries (the shifts are
small unless J$_H$ is large) and agrees\cite{f2} with Pai\cite{pai}. We show 
in Fig.3 the angle of canting for both $\theta_z$ and $\theta_{xy}$ deep
inside the G-phase at $x=0.98$. The angles in Fig.3 represent deviation from
180$^{0}$. There is hardly any canting in $\theta_z$ while 
in $\theta_{xy}$, there is no significant canting for low J$_H$ and it is
about 10$^o$ only for large J$_H$.
We note that experiments\cite{akimoto98,kajimoto99} have so far not been
able to detect any significant canting in A- and C-phases. Even in the
G-phase, certain systems appear to show little canting. 

The interactions (both intra- and inter-orbital) are treated in the mean-field
(MF) theory and a self-consistent calculation has been performed.
Self-consistency is achieved when all the averages $<\hat{n}_{i,\sigma,\alpha}>
$ and the ground state energy converge to within 0.01\%. Figs.1a,b and 2a,b
show the modifications in the magnetic phase diagram by the inter-orbital
($U^\prime$) interaction in the $x-J_{H}$ plane at
$J_{AF}S_{0}^{2}=0.05$ and in the $x-J_{AF}$ plane for $J_{H}S_{0}= 10$. 
Although we obtained the phase diagram for several values of $U^\prime$, in
Figs.1,2 we only give representative ones for demonstration.
On increasing $U^\prime$, the F-phase starts shrinking fast, the C-phase
gains in size while the G-phase remains almost unaltered. This is primarily
because of the enhanced orbital ordering in the A- and C-phases driven by
the inter-orbital repulsion and the low dimensional nature of the
DOS in these phases. As 
discussed earlier the AFM A and C phases are driven by orbital ordering
and in the presence of $U^\prime$, the one dimensional order leading
to the AF instabilty in the C-phase grows faster. Close to the $x=1$ end
the electron density is very low, there are almost no sites with both the
orbitals occupied and $U^\prime$ is therefore ineffective. At the other end,
however, the density is higher and the F phase has preferential
occupation of one spin at both the orbitals. Hence this phase is affected 
drastically by the inter-orbital repulsion.

It is known\cite{hotta00,dagrev} that at the level of MF theory the intra-
orbital repulsion $U$ between opposite spins mimics the effect of J$_H$. 
As we are
working with quite low densities (actual filling $\le$ 0.125), and the
relevant J$_H$ values being large, we find almost no observable effect of $U$
on the phase diagram (except for very low J$_H$ where again the changes are
small).

In the absence of interactions there is orbital ordering in both A- and
C-phases. The presence of U$^\prime$ enhances this ordering.
We calculate the orbital densities in both A- and C-phases in the
respective ground states and show the
results in Fig.4. In the figure, we have plotted actual orbital occupancies 
(the sum of the occupancies of the two orbitals will be $\frac{1-x}{4}$)
for different $U^\prime$. It is evident from the figure that
in the A-phase the  d$_{x^2-y^2}$ orbitals are predominantly occupied, while
in the C-phase the d$_{z^2}$ orbitals have higher occupancy. As $U^{\prime}$
increases, the orbital ordering is enhanced. This is shown in Fig.4 for three
values of $U^\prime$ in the A and C phases in their respective regions of
stability as a function of doping. Note that as $x$ increases (density of
electron decreases), the effect of $U^\prime$ on the orbital occupancies 
becomes less pronounced and the curves for different $U^\prime$ merge as
expected. We also show the
orbital occupancies as a function of $U^{\prime}$ in Fig.5 in the regions
where A- and C-phases are stable and the effect of U$^\prime$ is noticeable
in both the A- and C-phases. The orbital densities in C-phase attain their
saturation values by  $U^\prime \simeq 8$. Since we are interested in the 
region $x \ge 0.5$, we have not plotted orbital densities in A-phase beyond
$U^\prime=8$ -- above this value A-phase shifts below $x=0.5$ at J$_H=5$
(see Fig.1).

The present calculations produce results that agree with BK and Pai for
U=U$^\prime$=0. We are able to recover the G-phase at $x\simeq 1$ and
we also obtained the F phase for J$_{AF}\simeq  0$. On inclusion of
inter- and intra-orbital interactions, the topology of the phase diagram
remains the same. The C-phase grows at the expense of F phase while the G-phase
remains unaffected with increasing $\frac{U^\prime}{t}$. This scenario
is borne out in the bandwidth controlled
experiments of Akimoto et. al.\cite{akimoto98} and the schematic phase
diagram obtained by Kajimoto et. al\cite{kajimoto01}. Our results qualitatively
agree with the earlier work of Maezono et. al.\cite{maezono} as well. They 
included correlations in an MF treatement, but did not get the A-phase
close to $x\simeq 0.5$ observed experimentally. Although Fig.15 in Maezono et.
al.\cite{maez2} resembles (with a vanishing A-phase close to $x=0.5$)
our Fig.2, a comparison is quite difficult owing to the very different choice
of the parameters (it is also not possible to separate the effects of Coulomb
and exchange interactions in their work). In their Monte Carlo
treatment, Shen and Ting\cite{sheng} considered an effective
model and obtained a phase diagram. However, the C-phase in the region $0.6\le
x \le 0.9$ does not come out of their work. Hotta et. al.\cite{hotta00} have
compared exact diagonalization results in one dimension with MF theory
and found the agreement to be good.
Our MF calculations also suggest that the qualitative trends obtained
are in good agreement with the physically expected and experimentally observed
results in the manganites.

We note that the value of U$^\prime$ for which the F-phase disappears from
the region $x \ge 0.5$ in our calculation is about $\frac{U^\prime}{t}=12.$
Depending on the value of $t$, corresponding $U^\prime$ is between 1.8 - 3.6 eV.
This value is somewhat on the  lower side for the range of values available in
literature (the range varies between 3-10 eV)\cite{dagrev,coey,misra}. 
Although the available values are actually bare values and in phases like
F and A, they are bound to go
down owing to metallic screening (a treatment of which is beyond the
scope of this work) -- a problem faced in all theories of correlated systems
across a metal-insulator transition. In the foregoing, we have assumed
that increase in interactions is qualitatively equivalent to reduction of 
bandwidth, while in reality, the interactions play more complex roles in
addition to charge localization which are not included in our calculation. 

In conclusion, we have included orbital correlations in a degenerate double
exchange model proposed by Brink and Khomskii for the electron doped, 
intermediate bandwidth manganites. We observe from a generalized mean-field  
calculation that the phase diagram captures most of the qualitative features 
seen experimentally. The orbital orderings obtained are in good agreement
with experimental observations. It also agrees with the
trends observed across several manganites with changing bandwidths.
It would be interesting to include JT coupling and extended range Coulomb
terms in the model and observe their effects particularly close to the
$x=0.5$ region.
 
\vspace{0.4cm}

\noindent{\bf Acknowledgement} 
\vspace{0.3cm}

We acknowledge extensive discussions and clarifications at the initial stage
with G. V. Pai. We also acknowledge helpful discussions with S. K. Ghatak 
and R. Pandit. The work is supported by a grant from DST (India). TM 
acknowledges support from CSIR (India) through a fellowship.

\newpage
\center {\Large \bf Figure captions}
\vspace{0.5cm} 
\begin{itemize} 

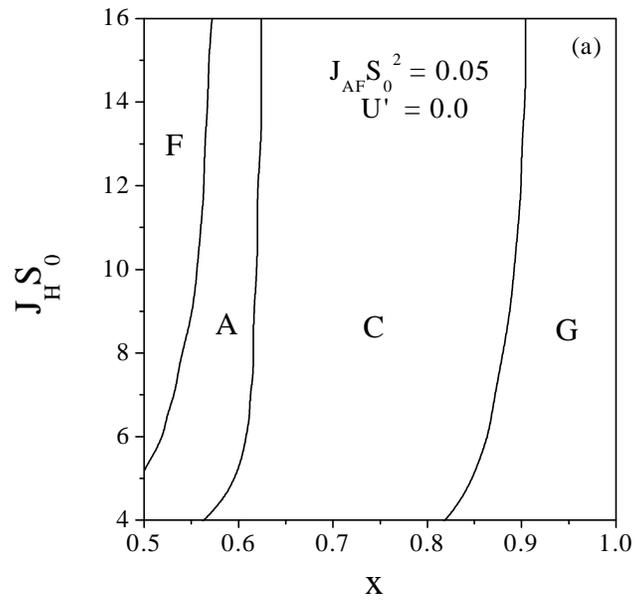
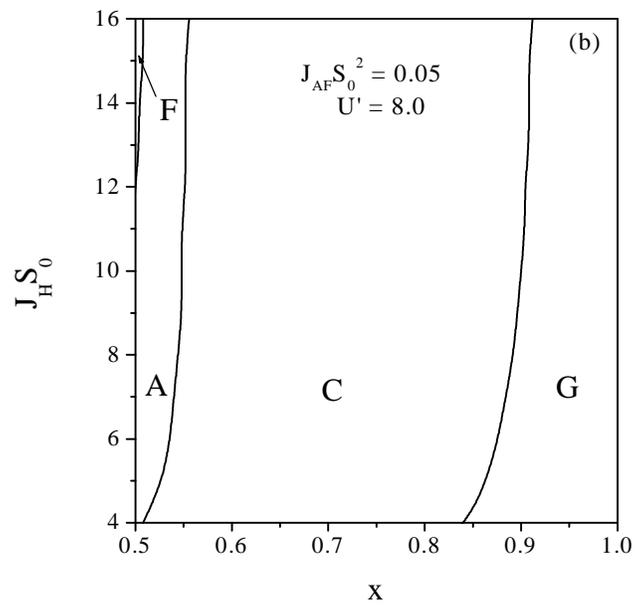
\item[Fig. 1.] Magnetic phase diagram in doping ($x$) - J$_H$ plane with (a)
$U^\prime=0$ and (b) $U^\prime=8.0$. All energies are measured in units of 
$t$.
 
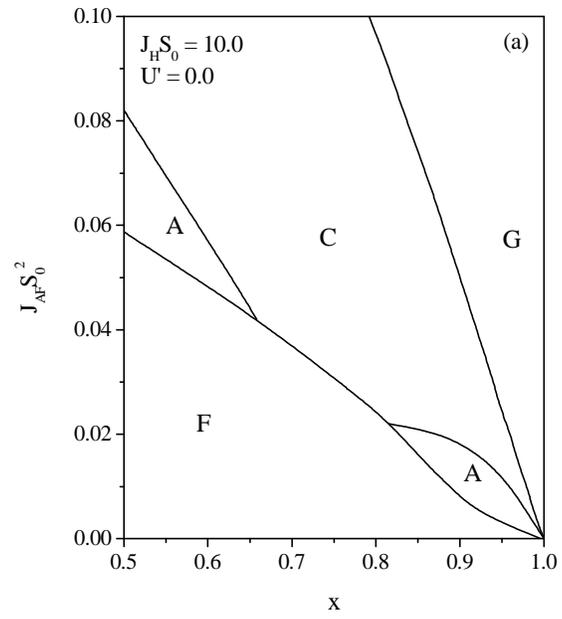
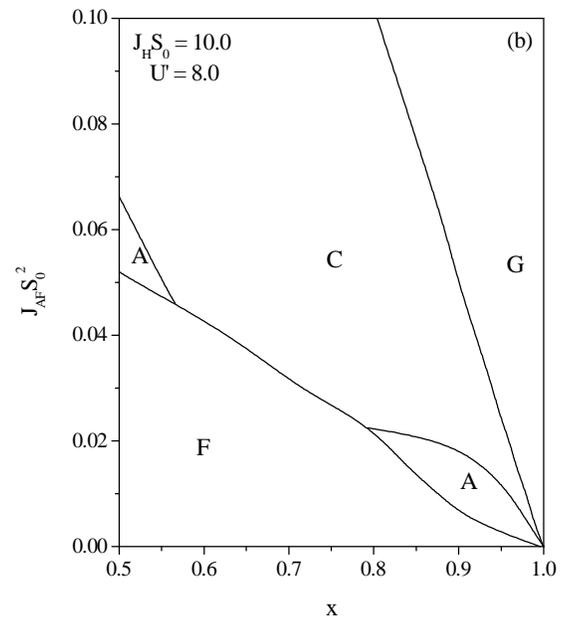
\item[Fig. 2.] Magnetic phase diagram in doping ($x$) - J$_{AF}$ plane with (a)
$U^\prime=0$ and (b) $U^\prime=8.0$. 

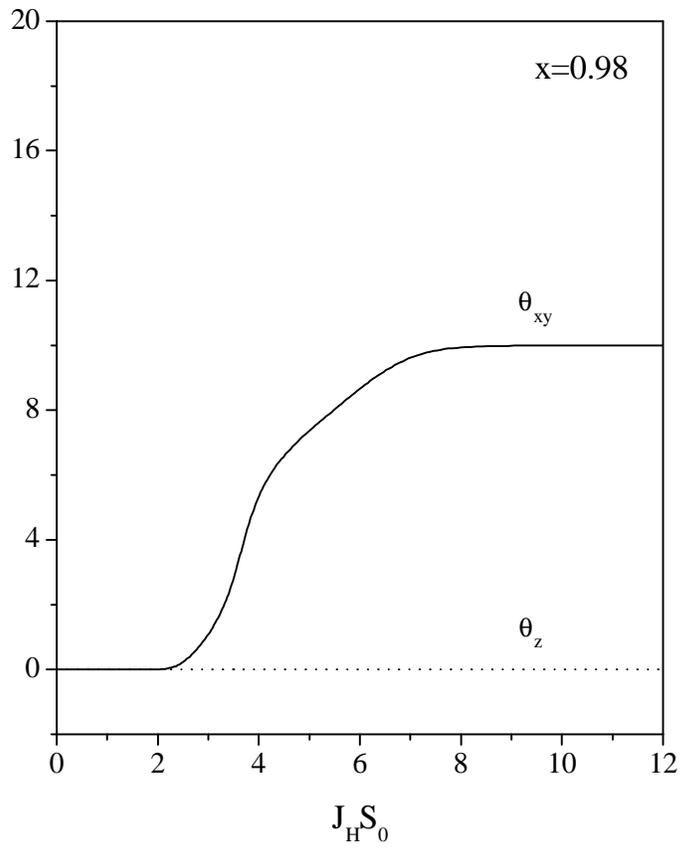
\item[Fig. 3.] Canting of the angles $\theta_{xy}$ and $\theta_{z}$ in degrees
as a function of J$_{H}$ (J$_{AF}S_{0}^{2} = 0.05$). 
 
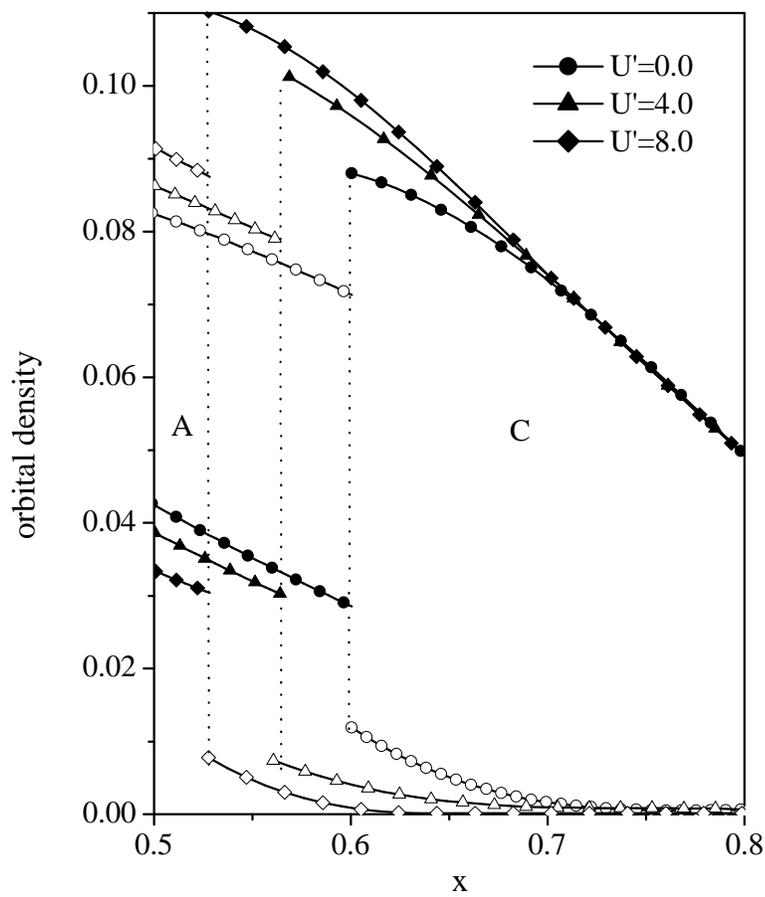
\item[Fig. 4.] Orbital densities as a function of doping $x$ for three values
of $U^\prime=0, 4, 8$. The filled symbols are for $d_{z^2}$ and open symbols for
$d_{x^2-y^2}$ orbitals. The vertical dotted lines represent the boundary
between A- and C-phases for different $U^\prime$. We choose J$_{H}S_{0}=5$
here in order to have stable A- and C-phases for a reasonable range of $x$
(see Fig.1) for all three $U^\prime$ values. J$_{AF}S_{0}^{2}$ was kept at
0.05. 

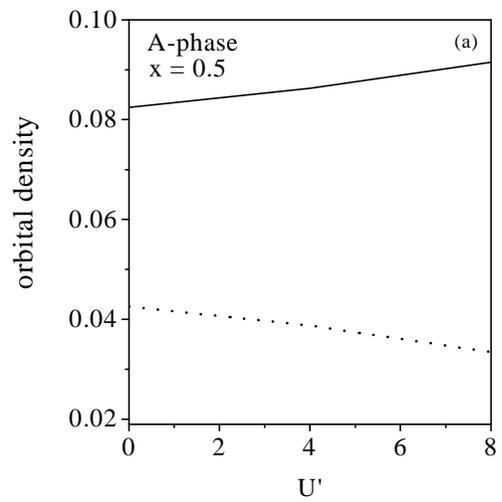
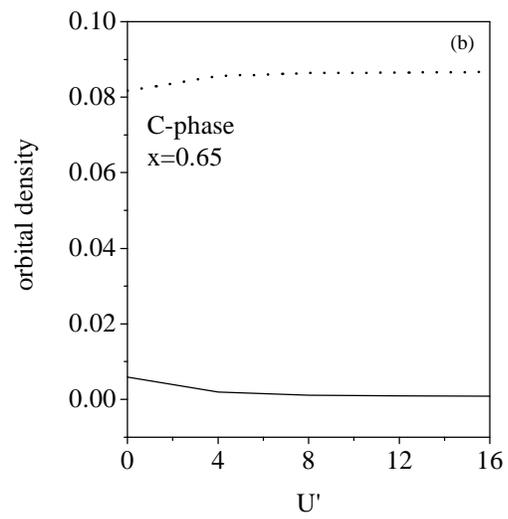
\item[Fig. 5.] Orbital density versus $U^\prime$ in (a) A-phase at $x=0.5$
and (b) C-phase at $x=0.65$. The dotted lines are for $d_{z^2}$ and solid
lines are for $d_{x^2-y^2}$ orbitals.  J$_{H}S_{0}$ and J$_{AF}S_{0}^{2}$ were
same as in Fig.4. 
\end{itemize} 

\vspace{0.4cm}

\newpage 
\begin{thebibliography}{999}
\bibitem{sdm} {\em Physics of Manganites}, edited by  T. A. Kaplan and S. D.
Mahanty (Kluwer Academic, New York, 1999). 
\bibitem{imada} M. Imada, A. Fujimori and Y. Tokura, Rev. Mod. Phys. {\bf 70},
1039 (1998).
\bibitem{tomi95} Y. Tomioka et. al,  Phys. Rev. Lett {\bf 74}, 5108 (1995).
\bibitem{kuwa99} H. Kuwahara et. al,  Phys. Rev. Lett {\bf 82}, 4316 (1999).
\bibitem{akimoto98} T. Akimoto et. al, Phys. Rev. B {\bf 57}, R5594 (1998).
\bibitem{kajimoto01} R. Kajimoto et. al., cond-mat/0110170. 
\bibitem{kajimoto99} R. Kajimoto et. al, Phys. Rev. B {\bf 60}, 9506 (1999).
\bibitem{yoshi} H. Yoshizawa et. al, Phys. Rev. B {\bf 58}, R571 (1998).
\bibitem{khom} J. Brink and D. Khomskii,  Phys. Rev. Lett {\bf 82}, 1016 (1999).
\bibitem{khom73} K. Kugel and D. Khomskii, Sov. Phys. JETP {\bf 37}, 725 (1973).
\bibitem{hotta00} T. Hotta, A. Malvezzi and  E. Dagotto, Phys. Rev. B
{\bf 62}, 9432 (2000). 
\bibitem{dagrev} E. Dagotto, T. Hotta and A. Moreo, Physics Reports, 
{\bf 344}, 1 (2001)..
\bibitem{coey} J. M. D. Coey et. al, Adv. in Physics {\bf 48}, 167 (1999).
\bibitem{misra} S. Misra, R. Pandit and S.  Satpathy, Phys. Rev. B {\bf 56}, 
2316 (1997); J. Phys. Cond. Matter, {\bf 11}, 8561 (1999). 
\bibitem{satpat} S.  Satpathy et. al, Phys. Rev. Lett. {\bf 76}, 960 (1996). 
\bibitem{pai} G. Venkateswara Pai, Phys. Rev. B {\bf 63}, 064431 (2001).
\bibitem{cuoco} Andrzej M. Oles, Mario Cuoco and N. B. Perkins, 
cond-mat/0012013.
\bibitem{comm_CO} The CE phase close to $x\le 0.5$ appears in some cases
in an extremely narrow region for the intermediate bandwidth
systems\cite{kajimoto01,akimoto98,kawano}. For 
(La$_{0.5}$Nd$_{0.5}$)$_{1-x}$Sr$_x$ MnO$_3$, the region of our interest
($z\simeq 0.4$ and $x\ge 0.5$) shows no charge ordering (see Figs. 4 \& 5
in\cite{akimoto98}). 
\bibitem{maezono} R. Maezono, S. Ishihara and N. Nagaosa, Phys. Rev. B.
{\bf 57}, R13993 (1998).
\bibitem{maez2} R. Maezono, S. Ishihara and N. Nagaosa, Phys. Rev. B.
{\bf 58}, 11583 (1998).
\bibitem{comm_BK} The phase identified as canted A-phase by BK is actually
a canted G-phase. This phase should be contrasted with the A-phase close
to $x=0.5$ end which is orbitally ordered.
\bibitem{sheng} L. Sheng and C. S. Ting, cond-mat/9812374. 
\bibitem{f2} In the choice of J$_H$S$_0$ in eqn.(3), there is an additional 
factor of 2 in ref.\cite{pai}. 
\bibitem{kawano} H. Kawano et. al., Phys. Rev. Lett. {\bf 78}, 4253 (1997).

\end{thebibliography}
\end{document}